\documentclass[aps,prc,floats,floatfix,showpacs,preprint,superscriptaddress,nofootinbib]{revtex4}

\usepackage{graphicx}
\def\be{\begin{equation}}
\def\ee{\end{equation}}

\def\e{\varepsilon}
{\catcode`\|=\active
  \gdef\Braket#1{\left<\mathcode`\|"8000\let|\bravert {#1}\right>}}
\newcommand{\bravert}{\egroup\,\vrule\,\bgroup}

\begin{document}

\title{Time-dependent density functional theory for X-ray near-edge
spectroscopy}

\author{G.F. Bertsch}
\affiliation{Institute for Nuclear Theory and Department of Physics,
University of Washington, Seattle, Washington}
\author{A. Lee }
\affiliation{Department of Physics,
University of Washington, Seattle, Washington}

\begin{abstract}

We derive a time-dependent density functional theory appropriate for
calculating the near-edge X-ray absorption spectrum in molecules
and condensed matter.  The basic assumption is to increase the space of
many-body wave functions from one Slater determinant to two.
The equations of motion derived from Dirac's variational principle
provide an exact solution for the linear response 
when the interaction Hamiltonian has only a core-electron
field.  The equations can be solved numerically nearly
as easily as the ordinary real-time time-dependent Kohn-Sham equations.
We carry out the solution under conditions that permit comparison
with the expected power-law behavior.  Our extracted power-law
exponents are similar to those derived by Nozi\`eres and DeDominicis, but 
are not in quantitative agreement.
We argue that our calculational method can be
readily generalized to density functionals that take into account
the more general electron-electron interactions that are needed for treating
dynamic effects such as plasmon excitations. 

\end{abstract}
\maketitle

\section{Introduction}

Time-dependent density functional theory (TDDFT) has proven to be a very 
useful tool for calculating the linear response of condensed matter to
electromagnetic probes.
The overall features of the dielectric function are reproduced quite 
well, and the agreement at zero frequency in insulators is often at 
the few percent level\footnote{It should be mentioned that there
are also well-known deficiencies in local density functionals,
namely band gaps are too high and excitons are missing from the
theory.}.  It also describes  the plasmon peaks
in the UV absorption spectrum and the corresponding energy loss
spectrum in inelastic electron scattering. 

The good experience in TDDFT in the optical regime encourages its
application to X-ray absorption, which has a rich near-edge structure
that is only partially described by single-electron physics. 
Indeed, some aspects of the X-ray response
are accessible to TDDFT\cite{pr09,sa10,lo12,le12a}.
However, other
aspects, particularly the dynamics of core-hole relaxation, are beyond
the scope of the theory as presently formulated. 
This is because the fundamental assumption
that the wave function be represented by a single Slater
determinant is too restrictive when the core orbital is
part of that determinant. The single-determinant theory can describe correlations between the
core hole and the valence electron but not correlation effects
within the valence space.  To overcome this deficiency, we propose here 
to extend the  
TDDFT from one determinant to two.  We derive the equations of motion
in Sect. II below.  We refer the reader to Ref. \cite{ma12} for an 
review of the foundations of TDDFT and its recent extensions.   
In Sect. III following we apply our extension to the well-known
Mahan-Nozi\`eres-De-Dominicis (MND) Hamiltonian.  A comprehensive
discussion of analytic and numerical methods to solve is given in the 
review by Ohtaka and Tanabe \cite{oh90}.  The equations we solve numerically
are exact for the MND Hamiltonian.  We will argue that they are much easier
than other methods to apply to DFT functionals which contain
electron-electron interactions.

\section{Extending TDDFT}

We view the TDDFT as an approximation to Hamiltonian many-body
theory taking the wave function as a single Slater determinant (SD).
The equations of motion may be derived from the
Dirac's variational principle
\be
\delta \,\,\int d t \Braket{ \Psi(t)| i{d \over d t} -
H|\Psi(t)} = 0
\ee
by varying the wave function $|\Psi\rangle$ in the space of
Slater determinants.  The resulting time-dependent Kohn-Sham
(KS) equations are then solved for the time-dependent wave function
$|\Psi(t)\rangle$.  For discussion of the action principle
in the context of TDDFT, see Ref. \cite{vi08,ru12}.

The linear response to an operator
$\cal O$ is obtained from the
time-dependent correlator $i \langle [{\cal O}(t),{\cal O}(0)]\rangle$
calculated from 
\be
\label{eq:R}
R(t) = \lim_{\lambda \rightarrow 0}{1\over \lambda}\langle \Psi_\lambda(t) 
| {\cal O} | \Psi_\lambda(t)\rangle.
\ee  
Here the initial state 
has been prepared by applying an impulsive field
$V(r,t) = \lambda {\cal O}\delta(t)$ to the KS ground state
$|\Psi_g\rangle$.  In linear order, this modifies the initial wave function to
\be
|\Psi_\lambda(0_+)\rangle =  (1-i\lambda{\cal O}) |\Psi_g\rangle.	
\ee
The resulting wave function is evolved by KS equations to determine the
matrix element in Eq. (2).  The connection of Eq. (2)
to the more familiar frequency-dependent response $S(\omega)$ is given
by
\be
\label{fourier}
 S(\omega) \equiv \sum_f \langle f | {\cal O} | 0\rangle^2
\delta(E_f-E_0 - \omega) ={1\over \pi}  \int_0^\infty d t R(t) \sin\omega t.
\ee

Since $\lambda$
is small in Eq. (\ref{eq:R}), the evolved wave function is still largely 
in the ground state with
only a small amplitude of excited states.  This is fatal for
calculating effects of the core hole excitation such as the relaxation
of the valence wave function in the presence of the core hole.
By considering separate determinants for the components of the wave function
with and without the core electron excitation, the correlations associated with
valence electrons can be treated as well as they are in the optical
response. There is 
no danger of violating the Fermi statistics because the two components
of the wave function are necessarily orthogonal.  We note that
multicomponent TDDFT has already been derived as a generalization of the 
Runge-Gross theorem and has been applied to a diatomic molecule\cite{le12b}.
The multi-determinant time-dependent theory is also well-known in the
literature \cite{be00,al07}.  However, it is typically 
based on a representation by particle-hole excitations of a single
determinant.  Our derivation and equations of motion are different,
resembling more the multi-determinant theory of nuclear excitations
proposed in Ref. \cite{pu13}.

Our starting point is the following 
ansatz for the variational wave function,
\be
\label{ansatz}
|\Psi\rangle = a_g c^\dagger_c|\Psi_g\rangle + a_c |\Psi_{c}\rangle 
=a_g c_c^\dagger \prod^{N_e}_\alpha c^\dagger_\alpha |\rangle + a_c
\prod^{N_e+1}_\beta c_\beta^\dagger |\rangle,
\ee
where $N_e$ is the number of active electrons in addition to the
core electron.
Here $g,c$ in the middle equality designate the determinants 
associated	 with the ground and 
core-excited wave functions,  respectively. The determinants are 
given more explicitly in the second equality, with $c^\dagger_\alpha,
c^\dagger_\beta$ creation operators in the valence band and $c^\dagger_c$
the creation operator for the core electron.  
The two sets of valence-band orbitals are expressed in terms of the valence-band
basis states $i$ as $c^\dagger_\alpha = \sum a_{\alpha i} c^\dagger_i
$ and $c^\dagger_\beta = \sum a_{\beta i} c^\dagger_i$.
Typically, the expansion coefficients $a_{\alpha i},a_{\beta i}$ satisfy the orthonormality
conditions
$\sum_i a^*_{\alpha i}a_{\alpha' i} = \delta_{\alpha\alpha'}$, etc.
The amplitudes of the two SD's, $a_g$ and $a_c$, should satisfy the
normalization condition
$ |a_g|^2 + |a_c|^2 = 1$.

The MND Hamiltonian has the form
\be
\hat H  = \hat H_v(c^\dagger_i,c_i) + 
\hat H_c(c^\dagger_i,c_i)c_c c_c^\dagger. 
\ee
The first term is the valence Hamiltonian to be constructed from the
corresponding Kohn-Sham density functional. The second term adds the
excitation energy of the core hole as well as its field acting on the
valence electrons.

The variation in Eq.~(1) is to be carried out with respect to
changes in the wave function $|\Psi\rangle$ that preserve its
character as a sum of two SD's.  In the single-determinant theory,
one takes the variational
derivatives of $|\Psi\rangle$ with respect to $a_{\alpha,i}$ treating
them as independent variables.  This results in the usual time-dependent
Kohn-Sham (KS) with the single-particle Hamiltonian given by\footnote{
The ordering of the operators in $\hat H_v$ is responsible for the
phase factor $s_{ij} = \pm 1$.}.
\be
\hat H_{KS} = \sum_{i,j} \langle\Psi | {\delta^2 \hat H_v \over \delta 
c^\dagger_i \delta c_j}
|\Psi \rangle c^\dagger_i s_{ij} c_j 
\ee 
However, as a consequence
of the overcompleteness of the variables,
the overall phase of the wave function no longer has
any physical meaning.  For example, if the orbitals are eigenstates of 
the KS Hamiltonian, the overall phase is 
$\exp(-i \sum_n^{N_e} \epsilon_n t)$ where $\e_n$ are the KS eigenvalues.  
The correct phase is 
$\exp(-i \langle \Psi |H | \Psi \rangle t)$; the two are only equal in the absence of
electron-electron interactions.   This phase plays no role in the 
single-determinant theory, but with two determinants it is crucial
to have correct relative phases.  

The proper procedure to apply Eq.~(1)
is require that the wave function variations 
$|\delta \Psi\rangle$ in derived equations
\be
 \Braket{ \delta\Psi| i{d \over d t} -
\hat H|\Psi(t)}=0.
\ee
are independent of each other. An orthogonal (and thus independent) set of 
wave functions may be defined 
by making use of Thouless's representation\cite{RS}  of the SD's.
The equations of motion are obtained by taking $|\delta \Psi\rangle$
as the set of 1-particle 1-hole excitations of the instantaneous SD,
\be
\label{delPsi}
|\delta \Psi\rangle \in |\alpha_p\alpha+h\rangle\equiv 
c^\dagger_{\alpha_p}c_{\alpha_h} |\Psi(t)\rangle.
\ee
in accordance with Thouless's theorem.  For a state of $N_e$
particles in a basis of dimension $N_b$, there are $N_e(N_b-N_e)$
particle-hole amplitudes to be determined compared to the $N_eN_b$
amplitudes in the representation Eq.~(\ref{ansatz}).  
However, the use of Eq.~(\ref{delPsi}) requires calculating both particle
and hole orbitals in an instantaneous basis, which is very costly in
carrying out the time evolution.  An easier 
way to avoid the phase introduced by the Kohn-Sham 
single-particle Hamiltonian is by projection.  The action of $\hat H_{KS}$
on the SD can be expressed in the instantaneous particle-hole basis as
\be
\hat H_{KS} | \Psi\rangle = E_{KS} |\Psi\rangle +
\sum_{\alpha_p,\alpha_h} v(\alpha_p,\alpha_h) |\alpha_p \alpha_h\rangle
\ee
where $E_{KS} = \langle \Psi | \hat H_{KS} | \Psi\rangle$.  The unwanted
first term can be removed in any basis simply by updating the wave function 
using the projected KS Hamiltonian $\hat H_{KS} - E_{KS}$.  
Thus, the single-particle orbitals
are calculated as usual, but the phase of the SD is corrected by
$\exp(+i \langle \hat H_{KS} \rangle \Delta t)$ at each time step.
For our numerical example below, the problem does not arise because
there is no electron-electron interaction in the valence space.

To summarize, we solve independently the time-dependent Kohn-Sham equations
for $|\Psi_g\rangle$ and $|\Psi_c\rangle$.  The two determinants
are coupled by the X-ray photon interaction,
\be
\hat H_x = v_x   (c^\dagger_x c_c+c^\dagger_c c_x), \,\,\,\,\,\,\,
c^\dagger_x = \sum_i f_i c^\dagger_i
\ee
with $f$  a form factor.  Varying with respect to $a_g,a_c$  one obtains
the $2\times 2$ matrix equation for these variables,
\be
\label{eq:ac}
i {d \over d t}\left(\begin{array}{c}
a_c\\a_g\end{array}\right) =  a_c \left(\begin{array}{cc}
\langle\Psi_c |\hat H - E_{KS,c}| \Psi_c \rangle &
v_x\langle \Psi_c|  \hat H_xc^\dagger_c  | \Psi_g\rangle\\
v_x\langle \Psi_g| c_c \hat H_x  | \Psi_c\rangle & 
\langle\Psi_g |\hat H -E_{KS,g} | \Psi_g \rangle \\
\end{array}\right) 
\left(\begin{array}{c}
a_c\\a_g\end{array}\right).
\ee
The hermiticity of $\hat H_x$ ensures that the normalization condition
remains satisfied during the course of the evolution. 
The off-diagonal matrix element in this equation is expressible as the
$N\!+\!1 \times N\!+\!1$ determinant
\be
\langle \Psi_c | c^\dagger_x | \Psi_g\rangle = 
\left| \begin{array}{ccc}
\langle\beta_1|\alpha_1\rangle & ... & \langle\beta_1|x\rangle\\
\langle\beta_2|\alpha_1\rangle & ... &  \\
    ...  & & \\
\langle \beta_{N+1}|\alpha_1\rangle & ... & \langle \beta_{N+1}|x\rangle\\
\end{array}
\right |.
\ee
While this determinant is well-known in the analytic theory of the
near-edge response\cite{oh90}, it is absent from the usual time-dependent 
Kohn-Sham theory based on a single Slater determinant.

To evaluate the linear response to the field of 
the X-ray photon,  
we start with the ground state wave function at time zero, 
$|\Psi_g\rangle$.  
We now perturb the
system by an impulsive X-ray field,  $\lambda \hat H_x \delta(t)$.
The immediate evolution introduces a small 
component of the core-excited
state into the wave function, 
\be
\label{eq:13}
|\Psi(0_+) \rangle = c^\dagger_c|\Psi_g(0)\rangle + 
i \lambda v_x|\Psi_c(0)\rangle,
\ee
where $|\Psi_c(0)\rangle =  c^\dagger_x |\Psi_g(0)\rangle$.
Eq. (\ref{eq:13}) has the required form as a sum of two determinants.
Each is evolved in time with its  own 
Kohn-Sham Hamiltonian.  Then the real-time response from Eq. (\ref{eq:R})
is 
\be
\label{eq:Rcg}
R(t)  = 2 v_x^2 {\rm Re}~\langle \Psi_c(t) | c^\dagger_x|\Psi_g(t)\rangle.
\ee
This can be easily Fourier-transformed by Eq. (\ref{fourier}	) to give the 
absorption spectrum.

Our procedure provides an exact solution for the linear response if
$\hat H_v$ and $\hat H_c$ are strictly one-body operators.  This is because
the intrinsically two-body part of the Hamiltonian does not make
two-particle excitations or entangle the two Slater determinants after
the initialization.

\section{Numerical calculations}

In this section we demonstrate the practicality of the method
as applied to the MND Hamiltonian.  The computer codes employed
here are available at Ref. \cite{computer}.  

We write the two terms in the Hamiltonian
as 
\be
\label{eq:Hv}
\hat H_v = {E_b\over N_b-1} \sum_{i=1}^{N_b} ( i - N_b/2) c^\dagger_i c_i
\ee
and
\be
\label{eq:Hc}
\hat H_c =  {v_c\over N_b}c^\dagger_x c_x, \,\,\,\, c^\dagger_x = 
\sum_i^{N_b} c_i^\dagger.
\ee
Here $E_b$ is the width of the band, and $N_b$ is thenumber of orbitals in
the band.  We shall express energies in units of $E_b$, and time in
units of $\hbar/E_b$.  We start with a half-filled band, taking the number
of valence electrons $N_e$ to be $N_e=N_b/2$.  We  present
calculations for the parameter sets listed in Table I.

As explained in the literature \cite{oh90}, the core-hole interaction
strength $v_c$ is not the most physically direct quantity determining the
near-edge response.  The effect of the Fermi-surface edge is 
more closely related to the shift of the single-particle orbital
energies due to the core hole.  Calling the shift $\Delta \e$, the
relevant quantity is 
\be
\Delta \e{d n \over d \e} = {\delta \over \pi}
\ee
where $dn/d\e$ is the density of orbital states at the Fermi level.
In the last equality, this is related to the scattering phase shift $\delta$
at the Fermi surface.  The values of $\delta$ associated
with the computed parameter sets are given in the last column of
Table I. 

The Green's function theory in Ref. \cite{no69}  for 
the time-dependent response 
decomposes it into two factors, the
overlap of Fermi sea determinants and the Green's function of the
electron that was promoted to the valence band.  
We write the overlap of the Fermi sea determinants as 
\be
    G(t) = \langle\Psi_g| e^{-i (\hat H_v + \hat H_c)t}|\Psi_g\rangle.
\ee
The main quantity of interest is the determinant in Eq. (\ref{eq:Rcg}) which
we call $g_c$, 
as 
\be 
g_c(t) = \langle\Psi_c(t)|c^\dagger_x\Psi_g(0)\rangle 
\ee
Nozi\`eres and De Dominicis decompose it into two factors, 
$g_c(t) = g(t)G(t)$. We will not make use of that
separation.

\subsection{Fermi sea evolution}
We first examine the Fermi sea overlap. To remove the phase of the
core-excited ground state, we will examine the quantity
\be
 G'(t) = 
e^{i \sum_\alpha \varepsilon_\alpha t}\langle\Psi_g| e^{-i (\hat H_v + \hat
H_c)t}|\Psi_g\rangle,
\ee  
where $\epsilon_\alpha$ are the Kohn-Sham eigenvalues of the ground-state
orbitals.
Figure 1 shows ${\rm Re}~ G'(t)$ for parameter set A of Table I.
\begin{figure}
\includegraphics [width = 11cm]{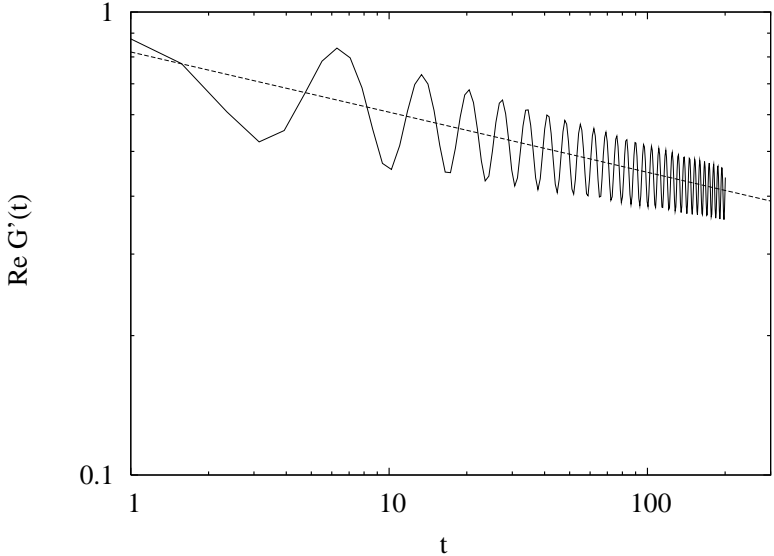}
\caption{\label{t-log} ${\rm Re}~
G'(t)$ as a function of time for parameter set A.  The line shows a visual power-law
fit, $ G'(t) \sim t^{-0.13}$.  
}
% ben:rt-xfs/xfs.G.py 256 128 -0.8  or do-it.sh
% log-fit.py 0.1 0.82 100 0.33 > t-log-fit.dat
% Gt_log.gnu
\end{figure}
It is plotted on a log-log scale to 
facilitate comparison with result of Ref. \cite{no69},
\be
\label{G-delta}
G(t) \sim (1/t)^{(\delta/\pi)^2}.
\ee
The predicted power-law dependence should be applicable over the time
domain starting from $t_0\sim 1/E_b$
and going to $t_1\sim d n /d \e = N_b/E_b$, the time
necessary to resolve individual orbitals in the band.  In our 
units the range is $(t_0,t_1) = (1,N_b)$. 
One notices immediately that $ G'(t)$ has a considerable
oscillatory component.  The oscillation has been found in other treatments
of the problem as well, eg. \cite[Eq. (3.4)]{oh86}.  As discussed in
Ref. \cite{oh90}, the oscillation may be attributed to
the deeply bound orbital at the bottom edge of the valence band.
\begin{table}[htb]
\begin{center}
\begin{tabular}{|c|cc|cc|}
\hline
case & $N_b$ & $N_e$ & $v_c$  & $\delta/\pi$  \\
\hline
A & 256 & 128 &  -0.8  & 0.38 \\
B & 8 & 4 &  -0.8    & 0.39 \\
C & 512 & 256 &  -0.8   & 0.38 \\
Z & 256 & 128 &  0    & 0 \\
%D & 256 & 128 &  0.0316   & 0.439 \\
\hline
\end{tabular}
\caption{Parameter sets for the Hamiltonian Eq. (\ref{eq:Hv},\ref{eq:Hc}). 
} 
\label{tableI} 
\end{center} 
\end{table} 
The line in the graph corresponds to the power law  $G(t) \sim
(1/t)^\gamma$ with $\gamma = 0.13$.  This is rather close to the predicted
power law derived in Ref. \cite{no69}, 
 $\gamma = (\delta/\pi)^2 \approx 0.14$.  

The spectral function associated with $ G'(t)$ is its Fourier
transform,
\be
G'(\omega) = \int_0^\infty d t e^{i \omega t}  G'(t).
\ee
\def\rGw{${\rm Re}~ G'(\omega)$}
Fig. \ref{fig:w} plots \rGw~ for parameter sets A and B.  For
the set A shown in the left-hand panel one can see the peaks associated with
individual states of the many-particle wave function.  The dimension of
the many-particle space is given by ${N_b \choose{N_e}} = 20$.  The ground state
is the peak on the far left, and 8-9 other states are 
visible in the plot.  The right-hand plot shows \rGw~
for parameter set A.  Here the individual states are so closely spaced
that one can see only smooth curves.  There are clearly two peaks
in the spectrum, one associated with the ground state and its 
low-energy excitations, and the other associated with a localized orbital
bound or
nearly bound orbital to the core hole.  In Fig. 3 we have replotted 
the \rGw~ ground-state peak on a log-log scale to make visible a power-law
dependence on $\omega$.  The expected range of validity for a power
law is within the interval $(\omega_0,\omega_1) = (1/N_b,1)$ in our units.    
The line in Fig. 3 shows the the power-law $\omega^{-1.13}$.  We can
see that it provides a reasonable fit in the range $(0.02,0.3)$ with
some oscillation at low frequency.  

\begin{figure}
\includegraphics [width = 8cm]{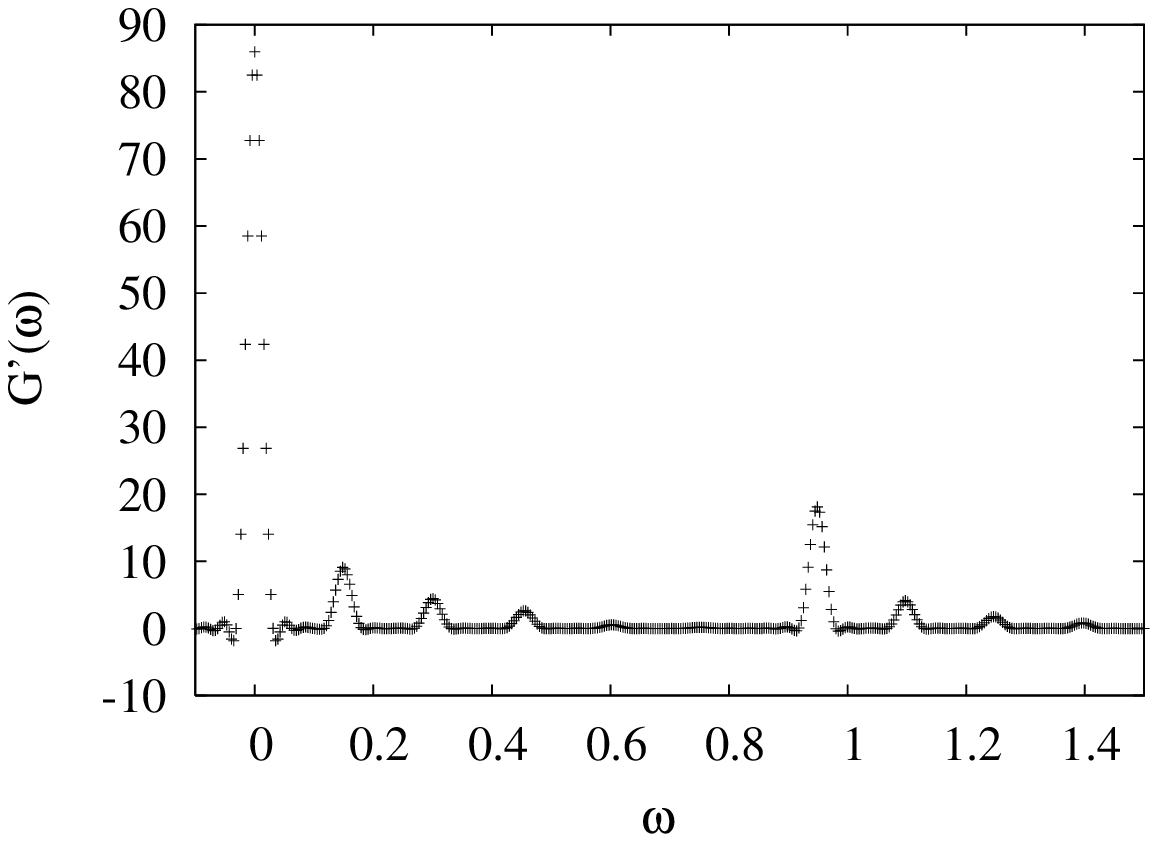}%
\includegraphics [width = 8cm]{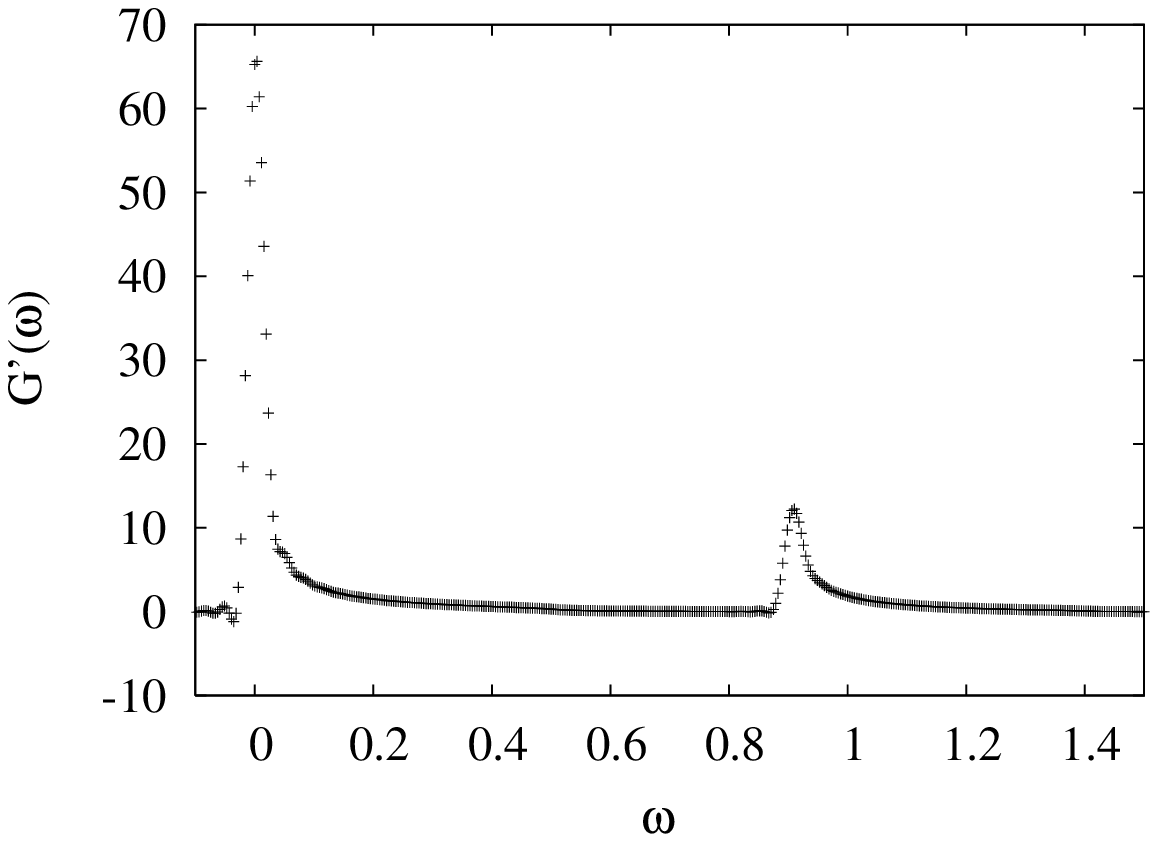}
\caption{
\label{fig:w} \rGw~as a function of $\omega$. Left-hand panel:
results for a low-dimensional system, parameter set B.
Right-hand panel: results for parameter set A.
}
%  rt-xfs/Gw.gnu --> Gw.eps; see do-it.py
\end{figure}
\begin{figure}
\includegraphics [width = 11cm]{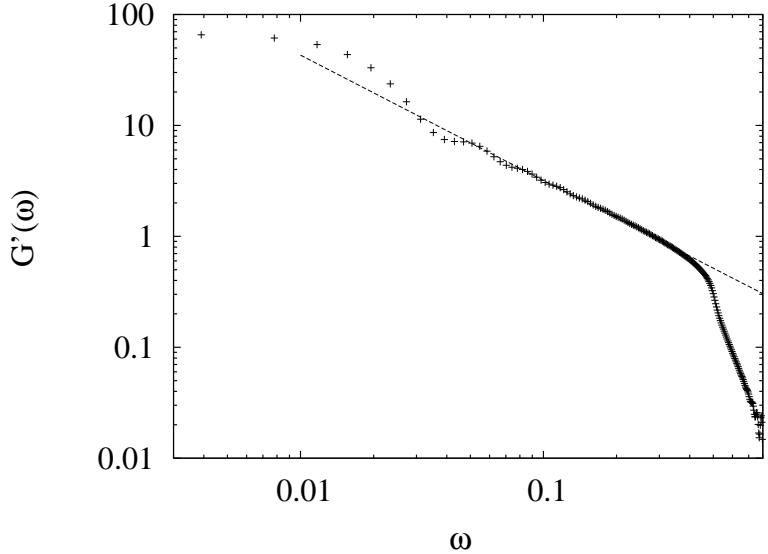}
\caption{\label{w-log} The same $ G'(\omega)$ as in Fig.2b, 
plotted on a log-log scale.  The line shows a visual power-law fit, 
$ G'(\omega) \sim
\omega^{-1.13}$.  
}
%ben:rt-xfs/Gw-logf.gnu
\end{figure}

\subsection{Inclusion of the core electron}

\begin{figure}
\includegraphics [width = 8cm]{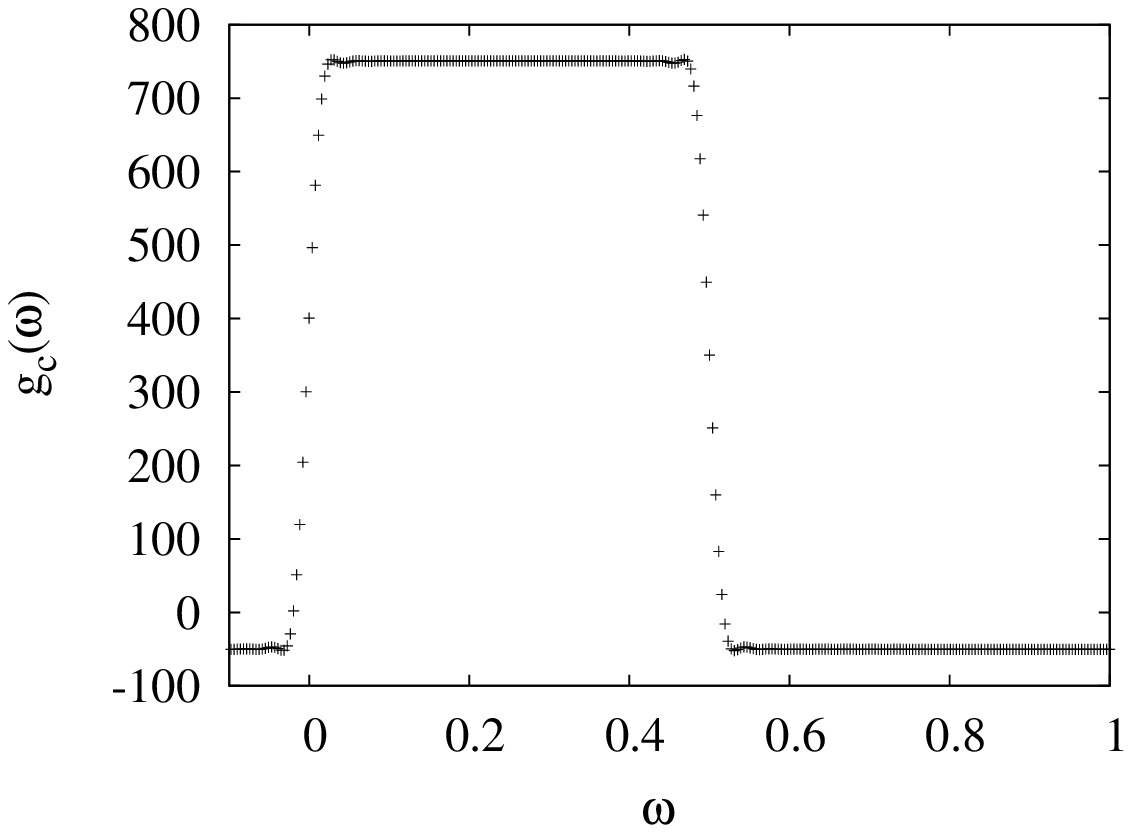}
\includegraphics [width = 8cm]{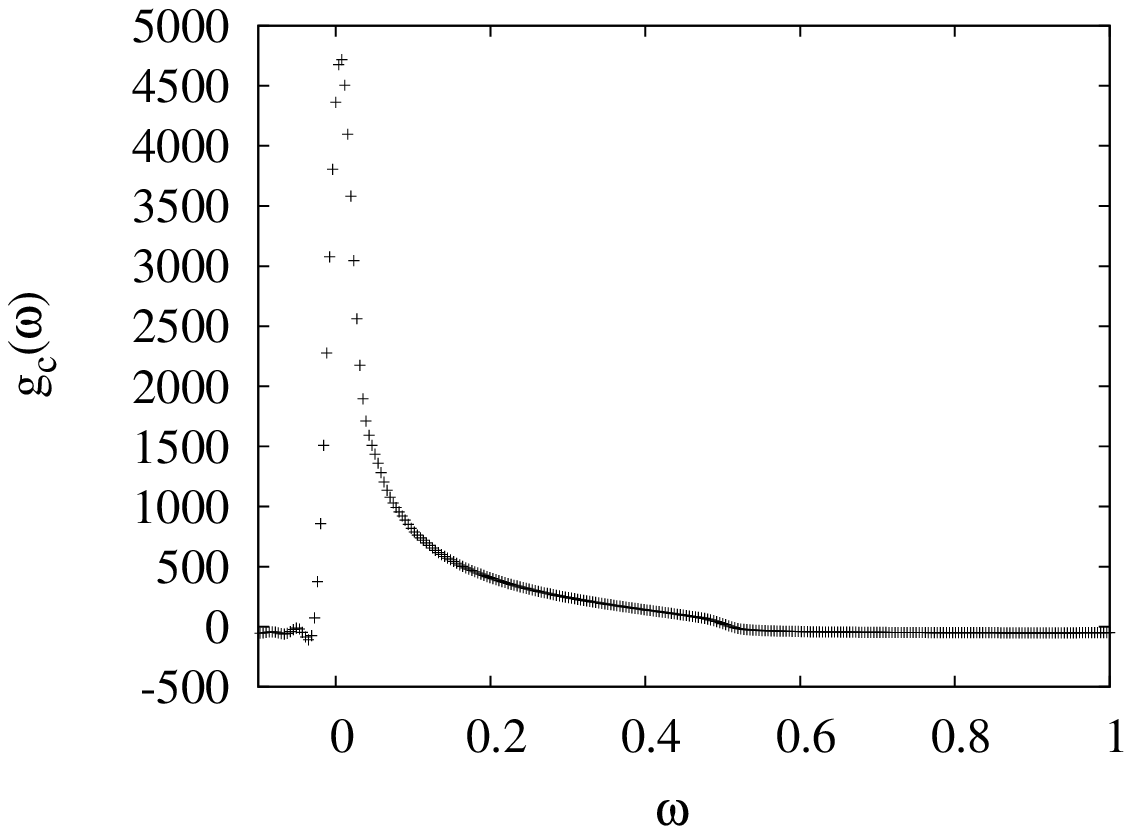}
\caption{\label{gw-0} Left panel: core-excited Green's function 
with no interaction, parameter set Z.  Right panel:  the same quantity
with parameter set A.  Note the difference in vertical scale.
}
% ben:rtgf2/rtgf.3.py 256 128 0.0
% ben:rtgf2/rtgf.3.py 256 128 -0.0316
\end{figure}

\begin{figure}
\includegraphics [width = 11cm]{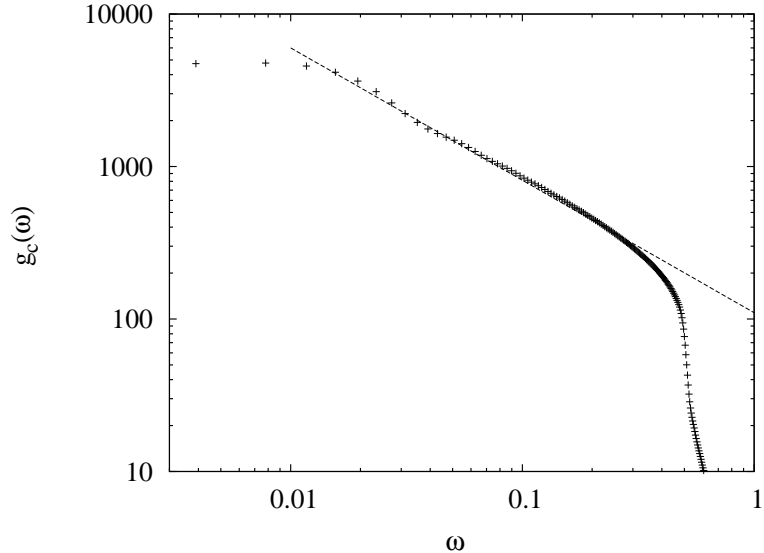}
\caption{\label{gw-log} The determinant  $ g_c(\omega)$, plotted on a log-log
scale.  The line shows an approximate power law fit, $g_c (\omega) \sim
\omega^{-0.85}$.  
}
\end{figure}

  We now examine the propagation of the
core-hole excited determinant with the core electron promoted to the valence
band.  The
number of electrons in the determinant is now $N_b/2+1$.
The initial wave function has equal amplitudes
for the x electron in all the unoccupied orbitals; it thus has the
same localization as the core-hole potential.  Just as a reminder
of the non-interacting physics, we show in Fig. (\ref{gw-0}) the
imaginary part of  $g_c(\omega)$ at $v_c=0$.  It is uniform
across the region of unoccupied orbitals,
with sharp edges at the Fermi surface and at the top of the band.
The right-hand panel shows the Green's function with parameter set A.
Note that the peak associated with a
hole at the bottom of the valence band is missing.  Evidently, the
electron added by $c^\dagger_x$ operator ensures with a high probability
that the hole is filled.   The results in the right-hand panel are plotted in
Fig. \ref{gw-log} on a log-log scale.  The line is a visual fit to power-law
behavior. Evidently, a power law gives a reasonable description over
the energy interval $0.02-0.3$.   According to the theory, the exponent is
determined by the phase shift according to the dependence
\footnote{This is for a single partial wave and spin projection.}.
\be
\label{eq:gamma}
\gamma =-2{\delta\over \pi}+{\delta^2\over \pi^2}.
\ee
Taking $\delta/\pi$ from Table I, the predicted value is $\gamma= -0.62$,
compared to $\gamma = -0.85$ from the fit.  

\begin{figure}
\includegraphics [width = 11cm]{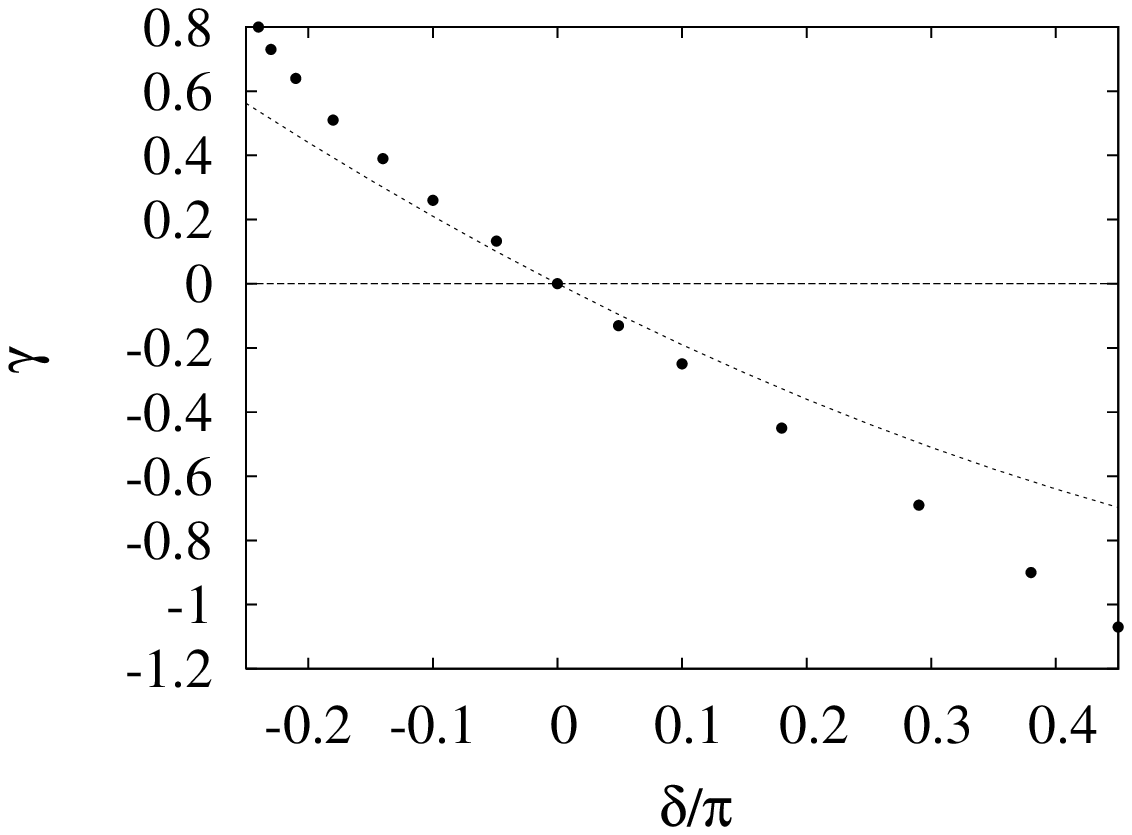}
\caption{\label{fig:gamma} Solid circles: the power-law 
exponent extracted from the extended TDDFT response for the space
$(N_b,N_e)=(256,128)$.  Dashed line: the analytic formula
Eq.~(\ref{eq:gamma}). The numerically computed   
exponent was extracted from 
the calculated $g_c(\omega)$ at $\omega=0.03$ and $0.2$ .
}
\end{figure}

The small disagreement we find here persists over 
a large
range of $\delta $.  Figure \ref{fig:gamma}. shows a comparison
over the range of $\delta$ accessible to the Hamiltonian.  We
see that the exponent is proportional to $\delta$ for small $\delta$
as in Eq.~(\ref{eq:gamma}), but the coefficient is somewhat higher.

\subsection{Other numerical treatments}

Various numerical treatments of the MND Hamiltonian have been discussed
and reviewed in Ref. \cite{oh90}.  The two main approaches are the
Green's function formulation\cite{sc77,gr77,pr01,ba80,oh86} and 
the formulation in terms of many-body
determinantal wave functions\cite{ko73,sw79,do80,fe80,ma80}.  The former 
requires constructing functions of at least two variables in the time or 
frequency domain, governed by equations that are nonlocal in those
variables.
In this respect, the real-time wave function 
method is much more efficient since there is only one time variable and
the equations to be solved are local in time.  An early numerical
work following the wave function approach is Ref. \cite{ko73}.  These
authors constructed the numerically exact solutions of the core-hole
excited Hamiltonian, and then used the ground state and one-particle excitations
of the Hamiltonian as the final states.  This
procedure of enumerating the eigenstates was also used in Refs. 
\cite{fe80}.  The wave function approach was also used in Ref. \cite{ma80},
and the determinant was evaluated in real time, as in our approach. 
However, the procedure adopted there was based on the formulation of
Ref. \cite{co71} which requires a matrix inversion.  The near-singularity
of the matrix apparently caused numerical difficulties that do not arise
in the real-time approach.  From a computational point of view, our
approach is closest to that used in Ref. \cite{sw79} and \cite{do80}.  We 
note that these authors found that the critical exponents of the
analytic Green's function treatment were only
in qualitative agreement with the numerical results outside of a very
small interval near $\omega=0$.

\section{Summary and outlook}

  We have derived an extension of time-dependent density
functional theory that contains at least some of the subtle many-particle
physics of X-ray near-edge absorption in metals.  Numerically, the real-time
theory is easy to carry out if the time-dependent electron-electron
interaction is neglected.  The absorption shows that the X-ray absorption
power-law behavior is in
qualitative agreement with the analytic results of Ref. 
\cite{ro69}, but not identical to them.  A similar conclusion was obtained
in Ref. \cite{sw79}.

Physically, the most
important effect of the many-electron physics is core-hole screening.  
There are several numerical calculations in the literature that follow
the Green's formalism of Ref. \cite{no69} and focus on this 
screening effect.   For the absorption spectra, a commonly used
approximation treats the system as fully relaxed in the presence of
the core hole.  Good fits can also be obtained under the assumption
that the dynamic screening reduces the core-hole effects
by a factor of two\cite{le12a}. That work also presented a real-time
dynamic calculation, but apparently used a
diagonal approximation to Eq. (\ref{eq:Rcg}).  
In any case, dynamic effects related to the core hole can be easily 
calculated in the real time method, so there is no reason to use any of 
these approximations.

So far we have not discussed 
the electron-electron interactions within the valence band.  They are
potentially important and are needed to treat the additional
screening associated with the plasmon degree of freedom.  Langreth 
has proposed a way to include plasmon effects in the Green's approach
\cite{la70} and it was applied with some success in Ref. \cite{gu12}.
However, it involves distinct calculations for the plasmon physics
and for the X-ray absorption.  In contrast, the real-time TDDFT 
provides a unified framework for including the electron-electron
interaction in the calculation. In the two-determinant theory
one can simply add the Coulomb field of the instantaneous charge density 
of $|\Psi_c\rangle$ to the field generated by $v_c$.  Of course,
the presence of the interaction requires considerably 
more computational effort.  The Coulomb field has to be calculated
at each time step.  Also, the overall phase of the $|\Psi_c\rangle$ has
to be computed using one of the methods discussed in Sect. II.  We believe
that the problem is still computationally quite tractable; we leave
the implementation to a future publication.

\section*{Acknowledgment}
We acknowledge discussions with J.J. Rehr, A. Bulgac, and K. Yabana.  We also
thank U. Mosel for bringing Ref.
\cite{pu13} to our attention.  This work was supported by the
National Science Foundation under Grant PHY-0835543 and by the 
(US) Department of Energy  under Grant No. DE-FG02-00ER41132.


\begin{thebibliography}{99}
\bibitem{pr09} M.~Prange, {\it et al.}, Phys. Rev. B {\bf 80} 155110 (2009).
\bibitem{sa10} A. Sakko, A. Rubio, M. Hakala, and K. Hamalainen, 
J. Chem. Phys. {\bf 133} 174111 (2010).
\bibitem{lo12} K. Lopata, B. Van Kuiken, M. Khalil, and N. Govind,
J. Chem. Theory Comput. {\bf 8}(9) 3284 (2012).
\bibitem{le12a} A. Lee, F. Vila, and J. Rehr, Phys. Rev. B {\bf 86}
115107 (2012).
\bibitem{ma12} M. Marque, N. Maitra, F. Nogueira, and E. Gross, eds.
``Fundamentals of time-dependent density-functional theory,'' 
Lecture Notes in Physics,  {\bf 837} (2012).
\bibitem{oh90} K. Ohtaka and Y. Tanabe, Rev. Mod. Phys. {\bf 62} 929 (1990).
\bibitem{vi08} G. Vignale, Phys. Rev. A {\bf 77} 062511 (2008).
\bibitem{ru12} M. Ruggenthaler and R. van Leeuwen, Ref. \cite{ma12}, pp. 206-210.
\bibitem{le12b} R. van Leeuwen and E. Gross, Ref. \cite{ma12}, p. 249.
\bibitem{be00} M. Beck, {\it et al.}, Phys. Rep. {\bf 324} 1 (2000)
\bibitem{al07} O. Alon, J. Chem. Phys. {\bf 127} 154103 (2007).
\bibitem{pu13} G. Puddu, arXiv:1208.0122v3,  (2013).
\bibitem{RS} P. Ring and P. Schuck, ``The Nuclear Many-Body Problem,''
(Springer, 1980).
\bibitem{no69} P.~Nozi\`eres and C.T.~De~Dominicis, Phys. Rev. {\bf 178} 1097
(1969).
\bibitem{pr01} T.~Privalov, R. Gel'mukhanov, and H.~Agren, Phys. Rev.
B{\bf64} 165115 (2001).
\bibitem{oh86} K. Ohtaka and Y. Tanabe, Phys. Rev. B {\bf 34} 3717 (1986).
\bibitem{sc77} K. Sch\"nhammer and O. Gunnarsson, Solid State Comm. {\bf 23}
691 (1977).
\bibitem{gr77} V. Grebennikov, Y. Babanov, and O. Sokolov, Phys. Status
Solidi B {\bf 80} 73 (1977).
\bibitem{ba80} U. von Barth and G. Grossmann, Physica Scripta {\bf 21}
580 (1980).
\bibitem{ko73} A. Kotani and Y. Toyozawa, J. Phys. Soc. Jpn. {\bf 35} 1082
(1973).
\bibitem{sw79} C. Swartz, J. Dow, and C. Flynn, Phys. Rev. Lett. {\bf 43} 
158 (1979).
\bibitem{do80} J.Dow and C. Flynn, J. Phys. C {\bf 13} 1341 (1980).
\bibitem{fe80} L. Feldkamp and L. Davis, Phys. Rev. B {\bf 22} 4994 (1980).
\bibitem{ma80} G. Mahan, Phys. Rev. B {\bf 21} 1421 (1980).
\bibitem{co71} M. Combescot and P. Nozi\`eres, J. Phys. (Paris) {\bf 32} 913
(1971).
\bibitem{ro69} B. Roulet, J. Gavoret, and P. Nozi\`eres, Phys. Rev. {\bf
178} 1072 (1969).
\bibitem{la70} D.C.~Langreth, Phys. Rev. B{\bf1} 471 (1970).
\bibitem{gu12} M. Guzzo, {\it et al.}, Eur. Phys. J. B {\bf85} 324 (2012).
\bibitem{computer}  See the URL
{\tt http://www.phys.washington.edu/users/bertsch/rt-xfs.tar} or
``Supplementary Material" accompanying the published
journal article. 
\end{thebibliography}
\end{document}